\documentclass[10pt, conference, a4paper]{IEEEtran}
\IEEEoverridecommandlockouts
\usepackage[numbers]{natbib}
\usepackage{url}
\usepackage{amsmath,amssymb,amsfonts}
\usepackage{algorithm}
\usepackage{algpseudocode}
\usepackage{graphicx}
\usepackage{multirow}
\usepackage{textcomp}
\usepackage{xcolor}
\usepackage{graphicx}
\def\BibTeX{{\rm B\kern-.05em{\sc i\kern-.025em b}\kern-.08em
    T\kern-.1667em\lower.7ex\hbox{E}\kern-.125emX}}
\begin{document}
\title{BeACONS: A Blockchain-enabled Authentication and Communications Network for Scalable IoV}

\author{Qi Shi, Jingyi Sun, Hanwei Fu, Peizhe Fu, Jiayuan Ma, Hao Xu and Erwu Liu
	% \thanks{Corresponding authors: Hao Xu and Erwu Liu.}
	\thanks{
		Q. Shi, J. Sun, H. Fu, P. Fu and J. Ma are with College of Electronic and Information Engineering, Tongji University, Shanghai 201804, PR China, E-mail:  \{qishi, 2252086, 2251039, 2252719, 2050971\}@tongji.edu.cn; 
	H. Xu and E. Liu are with College of Electronic and Information Engineering and Shanghai Engineering Research Center for Blockchain Applications And Services, Tongji University, Shanghai, China, E-mail: \{hao.xu, erwu.liu\}@ieee.org.
	}}

\maketitle

\begin{abstract}
This paper introduces a novel blockchain-enabled authentication and communications network for scalable Internet of Vehicles, which aims to bolster security and confidentiality, diminish communications latency, and reduce dependence on centralised infrastructures like Certificate Authorities and Public Key Infrastructures by leveraging Blockchain-enabled Domain Name Services and Blockchain-enabled Mutual Authentication. The proposed network is structured into a primary layer, consisting of Road Side Units and edge servers as servers of Blockchain-enabled Domain Name Services for managing inter-vehicle communications identities, and a sub-layer within each vehicle for intra-vehicle communications via the Blockchain-enabled Mutual Authentication Protocol. This design facilitates secure connections across vehicles by coordinating between the layers, significantly improving communications security and efficiency. This study also evaluates Road Side Unit availability against the random distribution of Road Side Units along the route of different vehicles. The proposed model presents a novel pathway towards a decentralised, secure, and efficient Internet of Vehicles ecosystem, contributing to the advancement of autonomous and trustworthy vehicular networks. 
\end{abstract}


\section{Introduction}
Internet of Vehicles (IoV) has become an emerging technology in recent years, especially in the context of smart transportation. IoV communications system can be divided into two main categories \citep{Taslimasa2023}: intra-vehicle and inter-vehicle. These two combined are also called Vehicle to Everything (V2X) communications. Intra-vehicle communications refer to all the communications between sensors, On-Board Units (OBUs) and Electronic Control Units (ECUs) inside the vehicle, while inter-vehicle communications refer to communications between vehicles (based on wireless communications modules such as OBUs) and other road entities such as Road Side Units (RSUs). Inter-vehicle communications include but are not limited to Vehicle to Vehicle (V2V), Vehicle to Infrastructure (V2I) and Vehicle to Pedestrian (V2P). 

However, IoV comes with not only exciting prospects but also various security risks and threats. Security issues in IoV include Denial-of-Service (DoS)/Distributed Denial-of-Service (DDoS), eavesdropping, impersonation, man-in-the-middle (MITM), spoofing and sybil attacks for inter-vehicle communications, and eavesdropping, masquerading, injection, DoS and message spoofing attacks for intra-vehicle communications \citep{Sharma2019}\citep{El-Rewini2020}. They undermine the confidentiality, integrity, privacy, authentication and availability of IoV. 

The absence of reliable communications identity authentication and secure confidential communications protocol provides the breeding ground for such issues. Meanwhile, due to IoV's nature of fast-changing connectivity and sensitivity to latency, it is also unwise to adopt excessively intricate, redundant security measures such as asymmetric session keys. Generally for IoV communications, connections should meet requirements in Table \ref{tab:Requirements of Targets}.

\begin{table}[htbp]
\caption{Requirements to be Satisfied by IoV Connections}
\begin{center}
\begin{tabular}{|l|l|}
\hline
\textbf{Requirements} & \textbf{Optimisation Target} \\
\cline{1-2} 
{Point-to-Point (P2P) Connection} & {Latency, Privacy}  \\ 
\hline
{Encrypted Session} & {Privacy, Confidentiality}  \\ 
\hline
{Secured Session Key Exchange} & {Privacy, Confidentiality, Authentication}  \\ 
\hline
{Zero-trust} & {Privacy, Authentication}  \\ 
\hline
{Certificateless} & {Availability, Privacy} \\
\hline
{Light-weight} & {Availability, Flexibility, Efficiency} \\
\hline
\end{tabular}
\label{tab:Requirements of Targets}
\end{center}
\end{table}

\begin{figure}[htbp]
\centerline{\includegraphics[width=0.5\textwidth]{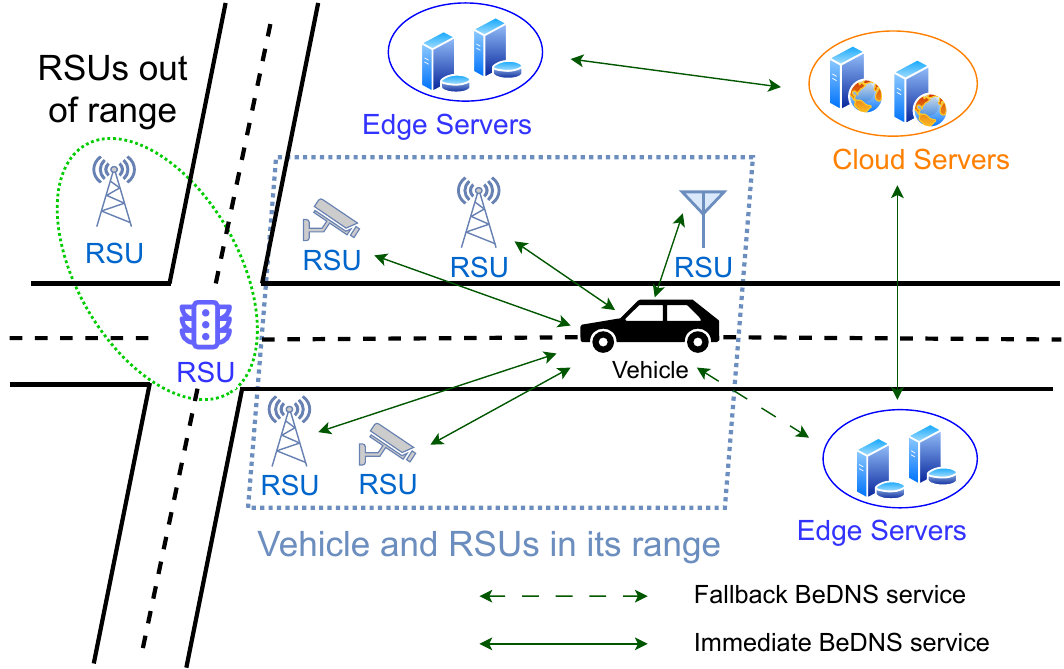}}
\caption{Overall structure of the primary layer of the proposed network}
\label{fig: PrimaryLayerStructure}
\end{figure}

In order to satisfy the requirements above, one type of communications identity management system is required for peer discovery and secure peer-to-peer routing of P2P connections. In IoV, servers and services of such a system should also meet requirements in Table \ref{tab:Requirements of Targets(2)}. 

\begin{table}[htbp]
\caption{Requirements to be Satisfied by IoV Communications Identity Management System}
\centering
\begin{tabular}{|l|l|l|}
\hline
\textbf{Object} & \textbf{Requirements} & \textbf{Optimisation Target} \\
\hline
\multirow{2}{*}{Servers} & Distributed to Edge & Latency, Availability, Coverage \\
\cline{2-3}
 & Decentralised & Integrity, Resilience \\
\hline
\multirow{3}{*}{Service} & Quick-to-manage & Immediacy, Latency \\
\cline{2-3}
 & Certificateless & Availability, Privacy \\
\cline{2-3}
 & Light-weight & Availability, Flexibility, Efficiency \\
\hline
\end{tabular}
\label{tab:Requirements of Targets(2)}
\end{table}

IoV is comprised of inter- and intra-vehicle networks, yet communications at these two different levels are not exactly segregated or arbitrarily mixed. Non-sensitive communications are required to be able to cross the two levels at the user's request, while information within the intra-vehicle network could disclose privacy or reveal security details if leaked to the inter-vehicle network. Therefore, the two types of networks should have separated identity management systems that conditionally exchange information in a controlled way. 

\subsection{Contributions}
\begin{itemize}
    \item This paper proposes the blockchain-enabled authentication and communications network for scalable IoV (BeACONS) that introduces Blockchain-enabled Mutual Authentication (BeMutual) into IoV as an encrypted, zero-trust, light-weight communications protocol for both inter- and intra-vehicle communications, which does not rely on centralised infrastructure such as Certificate Authority (CA) and Public Key Infrastructure (PKI). 
    \item The proposed BeACONS also introduces Blockchain-enabled Domain Name Services (BeDNS) into IoV as an immutable decentralised name service for communications identity management, which guarantees secure routing and in-time update of routing information. This paper also integrates BeDNS servers into RSUs to distribute services to the edge and assesses RSU-based BeDNS service availability. 
\end{itemize}

\section{Related Works}
\subsection{Certificateless Protocols}
\citet{Cui2018} proposed an efficient certificateless aggregate signature (CLAS) scheme without pairings for V2I communications. However, \citet{Kamil2019} proved that the CLAS scheme in \citep{Cui2018} is insecure against a polynomial time Type II adversary $\mathcal{A_{\text{2}}}$. Accordingly, authors of \citep{Kamil2019} proposed a refined CLAS scheme based on Elliptic Curve Cryptography (ECC) for vehicular ad hoc networks. Nevertheless, it still has weakness demonstrated in \citep{Zhao2020}: it is ineffective against the type I adversary $\mathcal{A_{\text{1}}}$'s and the type II adversary $\mathcal{A_{\text{2}}}$'s attacks. As of now, researches like \citet{Xie2023} \citet{Genc2023} not only focus on certificatelessness but also dig deep into the topic of conditional privacy preservation. \citet{Xu2021} proposed Blockchain-enabled Mutual Authentication (BeMutual) as a novel secure and privacy-preserving P2P communications protocol. Results show that BeMutual improves communications and computation overheads compared to the existing communications authentication protocols such as TLS 1.3 and IKEv2. 

\subsection{Decentralised Scheme Based on Blockchain for IoV}
Blockchain technology has been applied to many areas including radio access network (RAN) \citep{Ling2020}\citep{Wang2021}, Internet of Things (IoT) \cite{Cao2023a} and federated learning (FL) \cite{Cao2023b} and is pushing Internet to the era of Web3 \cite{Liu2023}. For Web3, \citet{Zhou2023} implemented Blockchain-enabled Domain Name Services (BeDNS) as a system of identity management for BeMutual. BeDNS maps complicated blockchain addresses to simpler domain names and blockchain addresses to their owners' network interface identifiers, which serves as a secure verification method in establishing BeMutual connections. For IoV, \citet{Jabbar2020} presented the blockchain-based Decentralized IoT Solution for Vehicles communications (DISV) which is effective against security, centralisation, and privacy leakage issues in V2X communications. For software-defined network (SDN) in IoV, \citet{Vishwakarma2022} introduced a lightweight blockchain-based security protocol for secure communications and storage in SDN-enabled IoV (LBSV), which is a permissioned blockchain network based on what the authors called the modified practical byzantine fault tolerance (mPBFT) consensus algorithm. 

\section{System Model}
From the previous works summarised above, this paper chooses BeMutual as the secure communications protocol for IoV connections, and BeDNS as the communications identity management system for inter-vehicle communications with or without intra-vehicle communications. For example, connections between two sensors on two different vehicles are inter-vehicle communications with intra-vehicle communications, while connections between two V2V communications modules on two different vehicles are purely inter-vehicle communications without intra-vehicle communications. Intra-vehicle communications within one vehicle, by contrast, are managed by wireless communications modules (such as OBUs) and don't require BeDNS. The proposed communications system therefore consists of two or more layers in terms of topology: the one-and-only primary layer that is in charge of inter-vehicle communications, and one sub-layer dedicated to intra-vehicle communications for each vehicle. 

\begin{figure*}[htbp]
\centerline{\includegraphics[scale=0.5]{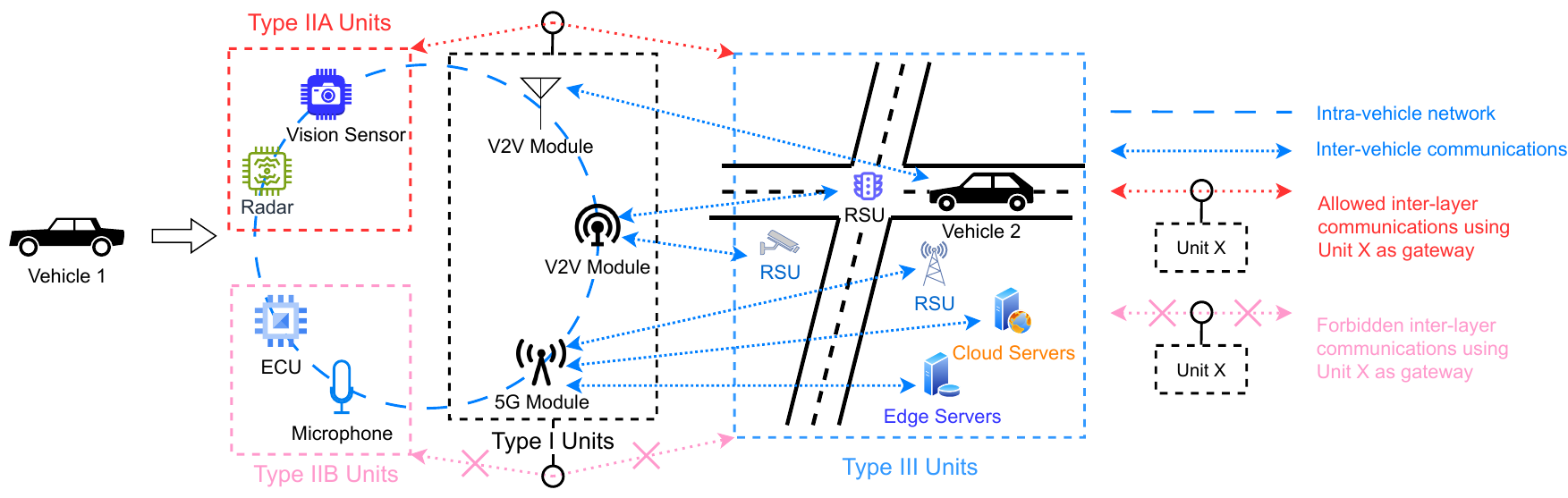}}
\caption{Overall structure of the sub-layer where intra- and inter-vehicle communications merge}
\label{fig: SubLayerStructure}
\end{figure*}

\subsection{The Overall Structure of the Primary Layer}
The primary layer is where inter-vehicle connections are established, which features BeMutual integrated into BeDNS that guarantees secure communications. The key elements of the primary layer are its participants and their characteristics. 
\subsubsection{Participants in the Primary Layer}
Participants in the primary layer are all the devices that engage in inter-vehicle communications. Every participant is considered an entity and has its unique Blockchain Address (BCADD). As shown in Fig. \ref{fig: PrimaryLayerStructure}, there are three categories of participants in the primary layer:  

\begin{itemize}
    \item \textbf{RSUs and edge servers}: They provide BeDNS services and are responsible for maintaining the blockchain of BeDNS. They receive requests for blockchain identity management including Bind, Update, Verify and Search \citep{Zhou2023} for mapping information between BCADDs, their corresponding network topological locations and network interface identifiers. 
    
    RSUs provide immediate services for vehicles nearby in normal scenarios, whilst edge servers serve as emergency fallbacks in case of no available RSUs. Also, edge servers provide immediate service for participants in the primary layer that are not designed to use RSU-based BeDNS service. The mapping information recorded in the blockchain of BeDNS is about relationships between BCADDs, their corresponding network topological locations and network interface identifiers. 
    
    \item \textbf{Clients}: In this model, clients are vehicles but may encompass more types of entities if required. Vehicles establish secure connections with participants in the primary layer over BeMutual, and interact with BeDNS servers to: 
    
    \begin{itemize}
        \item Create/regularly update their mapping information. 
        \item Verify/search for mapping information to be used in establishing BeMutual connections. 
    \end{itemize}
    
    \item \textbf{Cloud servers}: They are participants in the primary layer that provide cloud services, such as cloud computing servers that assist with autonomous driving, Over-the-air (OTA) update delivery servers and remote diagnosis servers. 
\end{itemize}

The constitution of each category is also detailed in Table \ref{tab:Constitutions and Examples}. 

\begin{table}[htbp]
\caption{Constitution of the three categories of participants in the primary layer}
\begin{center}
\begin{tabular}{|l|l|l|}
\hline
\textbf{Category}& \textbf{Constitution} & \textbf{Example}\\
\cline{1-3} 
\multirow{2}{*}{RSUs/edge servers} & {RSUs} & {Smart traffic lights} \\ \cline{2-3}
& Edge servers & Edge computing servers \\ \hline
{Clients} & {Vehicles and more} & Taxis, trucks, etc. \\
\hline
Cloud servers & Cloud service vendors & OTA update servers \\
\hline
\end{tabular}
\label{tab:Constitutions and Examples}
\end{center}
\end{table}

\subsubsection{Characteristics for Participants in the Primary Layer}
In order to ensure the best performance of the system, and also to defend RSUs from cyberattacks, participants in the primary layer have four characteristics in the context of IoV, which makes them a little different from the general BeDNS model. 
\begin{itemize}
    \item \textbf{BeMutual/BeDNS priorities}: To ensure that the whole system is lightweight and dedicated, and also to avoid compatibility issues, usage of BeMutual and BeDNS is restricted to safe driving and privacy-sensitive information. Whether a connection should use BeDNS and/or BeMutual depends on its pre-agreed level of criticality and is prescribed using built-in tools such as the routing table. 

    \item \textbf{Partial isolation of RSUs}: This model assumes that RSUs are linked over the Internet. In order to defend against DDoS attacks from the Internet, and also to ensure that these RSUs are IoV-dedicated, RSUs have been configured to receive service requests only from direct wireless links with nearby vehicles. This can be easily and reliably implemented because such links and RSU-to-Internet connections often use physically separated network interfaces. By contrast, edge servers are set to accept requests from the Internet. 

    \item \textbf{Limited service range of RSUs}: For RSUs that serve vehicles nearby, their service range is naturally restricted due to the limited coverage of wireless signals that transmit vehicle-to-RSU communications. As is shown in Fig. \ref{fig: PrimaryLayerStructure}, for an individual vehicle, RSU availability keeps changing as the vehicle constantly leaves or enters the service ranges of RSUs. During the available time slot (ATS) of an RSU, it provides BeDNS services for the vehicle. 

    \item \textbf{Access to Secret Keys (SKs)}: In principle, the SK from which a BCADD is derived should be kept to the owner of the entity the BCADD stands for. In the primary layer, SKs of the participants and their owners are listed in Table \ref{tab:Access to SKs in the primary layer}. 
\end{itemize}

\begin{table}[htbp]
\caption{Access to SKs in the primary layer}
\begin{center}
\begin{tabular}{|l|l|}
\hline
\textbf{Participant} & \textbf{SKs accessible to} \\
\hline
RSU/edge server & Operator/owner \\
\hline
\multirow{2}{*}{Client (vehicle)} & Driver/owner\\
\cline{2-2}
& Communications modules of the vehicle \\
\hline
Cloud server & Operator/owner \\
\hline
\end{tabular}
\label{tab:Access to SKs in the primary layer}
\end{center}
\end{table}

\subsection{The Overall Structure of the Sub-layer}
The sub-layer is where intra-vehicle connections over BeMutual are established without BeDNS. Each vehicle has its own unique, exclusive sub-layer. The key elements of the sub-layer are its units and functions. 

\begin{table*}[htbp]
\caption{Features and examples of the four types of units in a sub-layer}
\begin{center}
\renewcommand{\arraystretch}{1.2}
\begin{tabular}{|c|l|l|l|l|l|}
\hline
\textbf{Unit Type}& \textbf{In $V_1$} & \textbf{Mapping info.} &\textbf{Inter-vehicle comm.} &\textbf{Intra-vehicle comm.} & \textbf{Example}\\
\cline{1-6} 
I & Yes & Uploaded & Yes & Yes & OBUs, cellular mobile communications modules \\
\hline
IIA & Yes & Uploaded & No & Yes & Non-sensitive vision sensors  \\
\hline
IIB & Yes & Not uploaded & No & Yes & ECUs, in-vehicle voice recorders \\
\hline
III & No & Uploaded & Depends & Depends & RSUs, Type IIA sensors in another vehicle \\
\hline
\end{tabular}
\label{tab:Features and examples}
\end{center}
\end{table*}

\subsubsection{Units in the Sub-layer}
Units in a sub-layer are divided into four types based on their communications capabilities, shown in Fig. \ref{fig: SubLayerStructure}. All of the units have their own unique BCADDs. 
\begin{itemize}
    \item \textbf{Type I}: Units that are capable of both inter- and intra-vehicular communications. They do not share the vehicle's BCADD; Instead, each of them has a different BCADD and these BCADDs are uploaded to the blockchain of BeDNS as part of the mapping information. 
    \item \textbf{Type IIA}: Units that are capable of only intra-vehicle communications themselves but may get involved in inter-vehicle communications with the help of Type I units. They also have their mapping information uploaded to the blockchain of BeDNS. 
    \item \textbf{Type IIB}: Units that are capable of only intra-vehicle communications and are not allowed to engage in inter-vehicle communications because they could leak privacy or pose safety risks if connected to untrusted parties outside the vehicle. The blockchain contains no mapping information concerning Type IIB units. 
    \item \textbf{Type III}: Units that are typically not within the intra-vehicle network of this particular vehicle but are capable of connecting to Type I and/or Type IIA units through inter-layer interactions if required. Of course, they also have their BCADDs uploaded to the blockchain of BeDNS. 
\end{itemize}

Table \ref{tab:Features and examples} shows the main features as well as examples of the four types of units in a random vehicle $V_1$. 

\subsubsection{Functions of Units in the Sub-layer}
In order to preserve privacy, integrity and security of the sub-layer in a light-weight way, the sub-layer has three major functions. 

\begin{itemize}
    \item \textbf{Communications identity management}: In the sub-layer, Type I units also assume the role of communications identity management system, providing mapping information between BCADDs of Type I/IIA/IIB units and their network interface identifiers in the intra-vehicle network. 
    \item \textbf{Access to SKs}: In the sub-layer, the SK of each unit is kept to the unit itself. Of course, the driver/owner of the vehicle may keep a copy of these SKs for maintenance purposes, but normally the driver/owner doesn't need them just to drive and park. 
    \item \textbf{Conditional mapping}: Whether a unit in the sub-layer should upload its mapping information to the blockchain of BeDNS is decided by the unit itself, if the driver/owner does nothing with his copy of SKs. This is because apart from the driver/owner, only the unit has the right SK to create a valid signature that is required for the Bind and Update function of BeDNS. 
\end{itemize}

\subsection{Interactions in the Proposed Network}
Participants in the primary layer and units in the sub-layer(s) interact with one another to establish secure BeMutual connections. Their interactions in the proposed network can be divided into: 
\begin{itemize}
    \item \textbf{Interactions within the sub-layer}: Refer to the interactions conducted only in the sub-layer of a single vehicle, with only Type I, Type IIA and Type IIB units involved. BeDNS is not involved in these interactions. 
    \item \textbf{Interactions within the primary layer}: Refer to interactions in the primary layer that involve RSUs and edge servers, vehicles and cloud servers. To be more specific about the case of vehicles in reality, it's actually Type I units of the vehicles rather than a general model of indivisible vehicles that are involved in interactions in the primary layer. 
    \item \textbf{Inter-layer interactions}: Refer to interactions that combine the two sorts of interactions mentioned above, where RSUs, edge servers, vehicles, cloud servers and Type I/IIA/III units are involved. 
\end{itemize}

\section{Implementation}
To demonstrate the nature of BeACONS, three of its features are detailed: data in the blockchain of BeDNS, the dynamic feature of RSU availability and interactions in the network to establish BeMutual connections. 

\subsection{Data in the Blockchain of BeDNS}
There are three different classes of mapping information in the blockchain: 
\begin{itemize}
    \item \textbf{Class I}: Mapping information concerning only BCADDs of RSUs, edge servers and cloud servers in the primary layer and their corresponding network interface identifiers (ADD). It shows how to reach an RSU/edge server/cloud server in the primary layer. 
    \item \textbf{Class II}: Mapping information concerning only BCADDs of vehicles and their corresponding Type I units, which shows how to connect to a vehicle. It is assumed that one vehicle has multiple different Type I units, each of which has a unique BCADD. A label is added for each Type I unit to specifically describe its function so that other participants know exactly which to connect to in a particular circumstance. 
    \item \textbf{Class III}: Mapping information concerning BCADDs of vehicles in the primary layer, their corresponding Type I units and Type IIA units in the sub-layers. It shows how to reach a Type IIA unit in inter-layer communications. Like Class II, a label is also added for each Type IIA unit to specifically describe its function. 
\end{itemize}

Table \ref{tab:Data structure} shows their data structure. 

\begin{table*}[htbp]
\caption{Data structure of the three classes of mapping information in the blockchain of BeDNS}
\begin{center}
\renewcommand{\arraystretch}{1.2}
\begin{tabular}{|c|l|}
\hline
\textbf{Mapping info. Class} & \textbf{Data} \\
\hline
I & M\{$BCADD_P$, $ADD_P$\}, T, S(M, T, {$SK_P$}) \\
\hline
II & M\{$BCADD_P$, $BCADD_{Ui}$, $ADD_{Ui}$, $L_{Ui}$\}, T, S(M, T, {$SK_P$}, $SK_{Ui}$) \\
\hline
III & M\{$BCADD_P$, $BCADD_{Ui}$, $ADD_{Ui}$, $L_{Ui}$, $BCADD_{Uiia}$, $ADD_{Uiia}$, $L_{Uiia}$\}, T, S(M, T, $SK_P$, $SK_{Ui}$, $SK_{Uiia}$) \\
\hline
\end{tabular}
\end{center}
\footnotesize{"Participant" is denoted by $P$, "Type I unit" $Ui$, "Type IIA unit" $Uiia$, "Label" $L$, "Mapping" M, "Timestamp" T and "Signature" S. }
\label{tab:Data structure}
\end{table*}

\begin{figure*}[htbp]
\centerline{\includegraphics[scale=0.7]{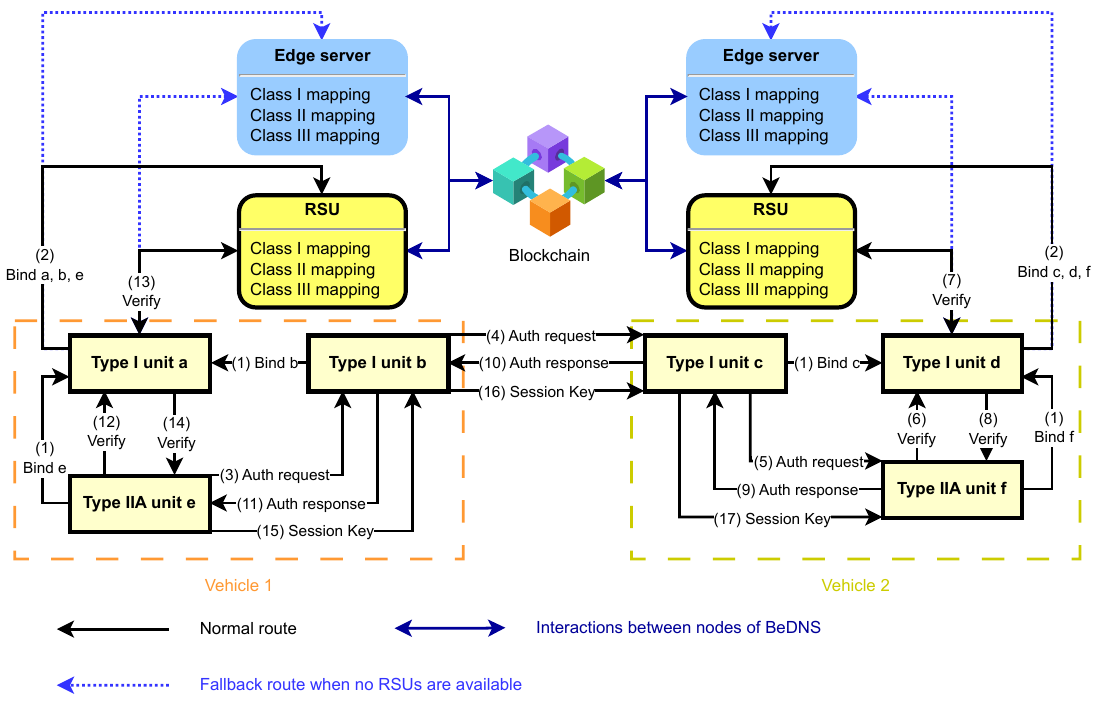}}
\caption{Establishment of inter-layer BeMutual session between two Type IIA units on different vehicles}
\label{fig: InterLayerInteractions}
\end{figure*}

\subsection{Dynamic feature of RSU availability}
As the vehicle keeps moving on the road, it constantly leaves or enters the service range of RSUs. For a group $G$ of RSUs $X_i$ serving one vehicle $V$, they absorb new RSUs coming into range and expel RSUs out of range. This process is carried out based on practical byzantine fault tolerance (PBFT) \cite{Castro1999}, where RSUs are nodes and the vehicle is the client. The primary node is denoted by $X_p$, and the RSU to be absorbed or expelled is denoted by $X_n$. 
\subsubsection{Absorption of an RSU}
The availability of an RSU is determined by the heartbeat message it regularly broadcasts. The procedure to absorb an RSU is demonstrated in Alg. \ref{alg: Abs}. 

\begin{algorithm}
\caption{Absorption of new RSU}\label{absorption}
\begin{algorithmic}[1]
\Procedure{Absorption}{}
    \If{reception of heartbeat of new RSU $X_n$ at $V$} 
        \State $V$ start BeMutual Authentication with $X_n$
        \If{BeMutual session is established}
            \State $V$ sends request of absorption of $X_n$ $Req_{Abs}$($X_n$) to $X_p$
            \If{PBFT consensus in $G$ on $Req_{Abs}$($X_n$) is reached}
                \State $N_{Xi} \gets N_{Xi} + 1$
                \State $X_i$ executes $Abs(X_n)$
                \State $X_i$ sends reply $Rpl_{Abs}$($X_n$) to $V$
                \If{reception of $\left(\frac{N_v+2}{3}\right)$ $Rpl_{Abs}$($X_n$) at $V$}
                    \State $V$ executes $Ack_{Abs}$($X_n$)
                    \State $N_{V} \gets N_{V} + 1$
                \EndIf
            \EndIf
        \EndIf
    \EndIf
\EndProcedure
\end{algorithmic}
\label{alg: Abs}
\end{algorithm}

In this procedure, $N_{Xi}$ is the number of nodes in $G$ recorded by $X_i$, while $N_V$ is the number of nodes in $G$ recorded by $V$. $Abs(X_n)$ is a procedure among $X_i$ to officially recognise $X_n$ as a member of $G$, which qualifies $X_n$ for receiving and sending messages of the PBFT procedure. In this process, the primary node $X_p$ does not change, thereby no view change. $Ack_{Abs}$($X_n$) is a procedure performed by $V$ that confirms absorption of $X_n$. The vehicle may interact with $X_n$ just like other $X_i$. 

\begin{figure*}[h]
    \centering
    \includegraphics[width=0.9\textwidth]{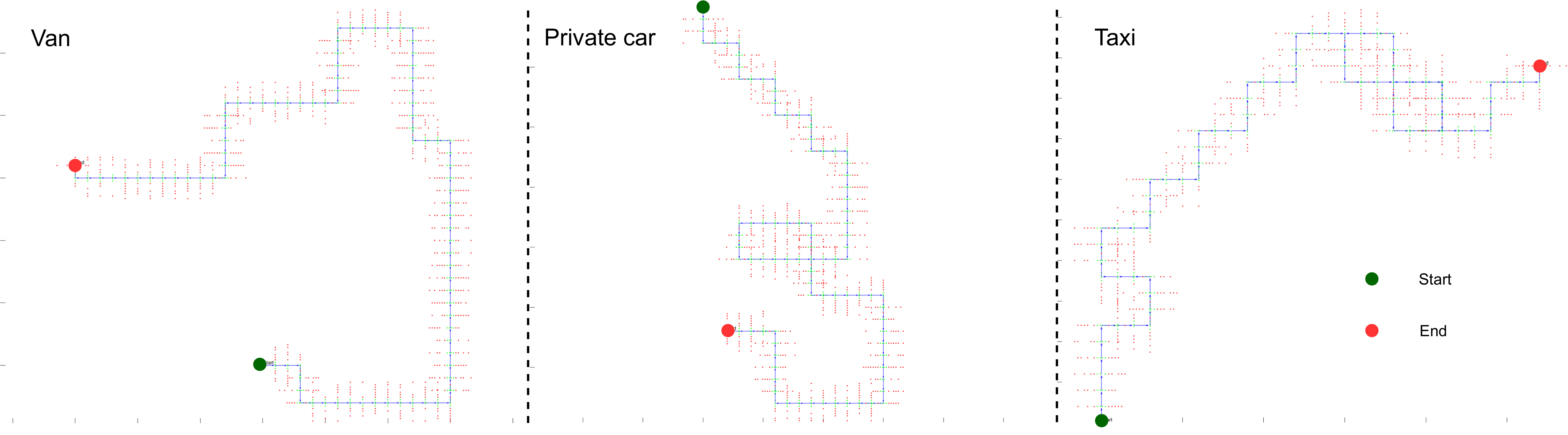}
    \caption{Representative routes of van (left), private car (centre) and taxi (right) which randomly travel on roads}
    \label{fig:avg}
\end{figure*}

\subsubsection{Expulsion of an RSU}
The procedure to expel an RSU is demonstrated in Alg. \ref{alg: Exp}, where $VC$($X_p$) is a special type of view change that automatically excludes $X_p$ from voting and being elected as the new primary node. After $VC$($X_p$) is executed, a new view is formed with a new $X_p$' and other $X_i$'. $Exp(X_n)$ is a procedure among $X_i$/$X_i$' to officially disqualify $X_n$ from $G$. $Ack_{Exp}$($X_n$) is a procedure performed by $V$ that confirms expulsion of $X_n$, after which $V$ no longer communicates with $X_n$. 

\begin{algorithm}
\caption{Expulsion of an RSU out of range}\label{expulsion}
\begin{algorithmic}[1]
\Procedure{Expulsion}{}
    \If{$V$ loses heartbeat of RSU $X_n$} 
        \State $V$ broadcasts to $X_i$ request of expulsion of $X_n$ $Req_{Exp}$($X_n$)
        \If{PBFT consensus in G on $Req_{Exp}$($X_n$) is reached}
            \If{$X_n$ is $X_p$}
                \State view change $VC$($X_p$) is executed
                \State $X_i'$ executes $Exp(X_n)$
                \State $N_{Xi'} \gets N_{Xi'} - 1$
                \State $X_i'$ sends reply $Rpl_{Exp}$($X_n$) to $V$
            \Else
                \State $X_i$ executes $Exp(X_n)$
                \State $N_{Xi} \gets N_{Xi} - 1$
                \State $X_i$ sends reply $Rpl_{Exp}$($X_n$) to $V$
            \EndIf
            \If{reception of $\left(\frac{N_v+1}{3}\right)$ $Rpl_{Exp}$($X_n$) at $V$}
                    \State $V$ executes $Ack_{Exp}$($X_n$)
                    \State $N_{V} \gets N_{V} - 1$
            \EndIf
        \EndIf
    \EndIf
\EndProcedure
\end{algorithmic}
\label{alg: Exp}
\end{algorithm}

\subsection{Interactions between Participants and Units}
Participants and units interact with one another to establish secure BeMutual connections. For the most general part, the establishment of an inter-layer BeMutual session is shown in Fig. \ref{fig: InterLayerInteractions}. In this process, Type I units serve as gateways to relay transmissions between Type IIA units and Type III units. Other kinds of interactions are simplified versions of the interactions in Fig. \ref{fig: InterLayerInteractions}. 

\section{Results}
For communications between RSUs and vehicles, ATS $t_1$ of RSUs must be long enough for BeDNS and other services/interactions to be effective, otherwise vehicle-to-RSU connections could be severed before a full cycle of service/interaction is complete. To evaluate the average number of RSUs with ATS $t_1$ longer than required (effective RSUs) along the way, an experiment is conducted. 

\subsection{Analysis}
The experiment features point motion along a polyline to simulate vehicle movement on the roadway. A Poisson distribution is used to model the probability of RSUs distributed along the perpendicular direction of the road, and the service range of each RSU is described as a circle centred at the location of the RSU (service circle). 

In reality, vehicles make turns randomly rather than keep driving along a straight line, which could affect ATS $t_1$ of RSUs situated near intersections. To simulate such scenarios, checkpoints are arranged at regular time intervals where the vehicle has a probability of turning or driving straight ahead. The exact figure of the probability of turning is determined by the vehicle type (taxi, truck, etc.). Based on every choice the vehicle makes at every checkpoint, it forms a random route along which it communicates with nearby RSUs. Random formation of a route is repeated 50 times to calculate the average number of effective RSUs for different types of vehicles. 

\begin{table}[h]
\centering
\caption{Simulation parameter settings}
\begin{tabular}{|l|l|}
\hline
Parameter                       & Value \\
\hline
Vehicle speed                   & \(v=1\) \\
Total travel time of vehicle                  & Time steps = 3600 \\
ATS                             & \(t_1=5\) \\
RSU density                     & \(\lambda_{\text{max}}=1\) \\
Interval between checkpoints    & Time steps = 20 \\
Radius of service circle               & \(r=7\) \\
Probability of turning & Depends on the type of vehicle \\
\hline
\end{tabular}
\label{tab:parameter settings}
\end{table}

\subsection{Parameters and Results}
Factors impacting the average number of effective RSUs have been identified and configured as shown in Table \ref{tab:parameter settings}. For taxis, private cars and vans which have different chances of turning, the average numbers of effective RSUs are listed in Table \ref{tab:Avg}. A representative route of each type of vehicle is shown in Fig. \ref{fig:avg}. 

\begin{table}[htpb]
    \centering
    \caption{Average number of effective RSUs against chance of turning}
    \begin{tabular}{|c|c|c|}
        \hline
        \textbf{Vehicle Type} & \textbf{Probability of turning} & \textbf{Avg.} \\
        \hline
        Van & 40\% & 4.26242 \\
        \hline
        Private car & 60\% & 4.844828 \\
        \hline
        Taxi & 80\% & 5.376702 \\
        \hline
    \end{tabular}
    \label{tab:Avg}
\end{table}

\section{Conclusion}
In this paper, a novel Blockchain-enabled Authentication and Communications Network for scalable IoV (BeACONS) is introduced. By decentralising identity management and authentication, BeACONS prevents common security threats such as eavesdropping and spoofing, thereby enhancing the confidentiality, integrity, and availability of communications for IoV. BeACONS features the integration of an identity management system into RSUs, and simulation results of RSU availability to nearby vehicles are provided. BeACONS provides a novel methodology for the development of secure and efficient IoV communications systems.

\renewcommand{\bibfont}{\footnotesize}
\bibliographystyle{unsrtnat}
\bibliography{export}

\end{document}